\documentclass[10pt,twocolumn]{IEEEtran}
\def\BibTeX{{\rm B\kern-.05em{\sc i\kern-.025em b}\kern-.08em
T\kern-.1667em\lower.7ex\hbox{E}\kern-.125emX}}

\usepackage{slashbox}
\usepackage{latexsym}
\usepackage{epsfig}
\usepackage{amsmath}
\usepackage{multicol}
\usepackage{amssymb}
\usepackage{subfigure}

\def\E{{\rm E}}
\def\C{{\rm C}}

\title{Coordinated Transmissions to Direct and Relayed Users in Wireless Cellular Systems}
\author{\begin{tabular}{c}
Chan Dai Truyen Thai, Petar Popovski, Megumi Kaneko and Elisabeth de Carvalho\\
\end{tabular}}

\begin{document}
\maketitle
\thispagestyle{empty}

\begin{abstract}

The ideas of wireless network coding at the physical layer promise 
high throughput gains in wireless systems with relays and multi--way traffic flows. This gain can be ascribed to two principles:
(1) joint transmission of multiple communication flows and (2) usage of \emph{a priori} information
to cancel the interference. In this paper we use these principles to devise new transmission schemes in 
wireless cellular systems that feature both users served directly by the base stations (direct users) and 
users served through relays (relayed users). We present four different schemes for \emph{coordinated transmission} 
of uplink and downlink traffic in which one direct and one relayed user are served. These schemes 
are then used as building blocks in multi--user scenarios, where we present several 
schemes for scheduling pairs of users for coordinated transmissions. The optimal scheme
involves exhaustive search of the best user pair in terms of overall rate. We propose several 
suboptimal scheduling schemes, which perform closely to the optimal scheme. The
numerical results show a substantial increase in the system--level rate with respect to 
the systems with non--coordinated transmissions. 
\end{abstract}

\section{Introduction}
Recently there have been extensive studies on cooperative, relay--based transmission schemes for extending cellular coverage or increasing diversity. Several basic relaying transmission techniques have been introduced, such as amplify-and-forward (AF) \cite{aaf1, aaf2}, decode-and-forward (DF) \cite{daf1, daf2} and compress-and-forward (CF) \cite{caf1}. These transmission techniques have been applied in one-, two- or multi-way relaying scenarios. 

In particular, two--way relaying scenarios \cite{anc, anc2, imperfect} have attracted a lot of attention, since it has been demonstrated that in these scenarios one can apply techniques based on network coding in order to obtain a significant throughput gain. There are two basic principles used in designing throughput--efficient schemes with wireless network coding: 
\begin{enumerate} 
\item \emph{Aggregation of communication flows}: instead of transmitting each flow independently, the principle of network coding is used in which flows are sent/processed jointly; 
\item \emph{Intentional cancellable interference}: in analog network coding, flows are allowed to interfere, knowing \emph{a priori} that the interference can be cancelled by the destination. 
\end{enumerate}
The motivation for this work was to generalize the two basic principles from above and devise novel transmission schemes in multi-user scenarios. 
We consider scenarios based on cellular networks with relays, where direct and relayed users are served in uplink/downlink.

Assume for example that a direct user wants to send a packet to the Base Station (BS), while the BS has a packet to send to a relayed user. In a conventional cellular system, these packets are sent over separate UL/DL phases. Instead, the BS may first send the packet which is received at the Relay Station (RS). While the RS forwards this packet to its intended relayed user, the direct user sends its packet to the BS, thus saving the required transmission time compared to the conventional method. We term such a scheme \emph{coordinated direct/relay (CDR)} transmission scheme. Transmission schemes that are related to some of the schemes proposed in this paper have appeared before in the literature~\cite{cites3, cites4}, or to relayed users \cite{nice}. However, in this paper we have used the principles described above to generalize the transmission schemes to in total four transmission schemes, which represent a superset of the existing schemes. As in the case of wireless network coding, our schemes take advantage of the combining of uplink and downlink traffic flows. Furthermore, we consider multi--user ($>2$) scenarios, in which the proposed CDR schemes are used as building blocks for creating novel scheduling schemes. We consider the rate--optimal scheme, which requires exhaustive search across the pairs of users and is complex. Therefore, we propose several suboptimal schemes and the results show that they perform closely to the optimal one, while all the proposed schemes show significant rate gains with respect to the reference system in which the CDR schemes are not employed. 

This paper is organized as follows. Section \ref{system model} describes the system model for two-user and multi-user networks. The two-user and multi-user schemes are described and analyzed in Section \ref{two-user schemes} and \ref{multi-user schemes}, respectively. Section \ref{numerical results} presents the numerical results and Section \ref{conclusion} concludes the paper. 

\section{System Model} 
\label{system model}
\begin{figure}
\centering
\includegraphics[width=0.6\columnwidth]{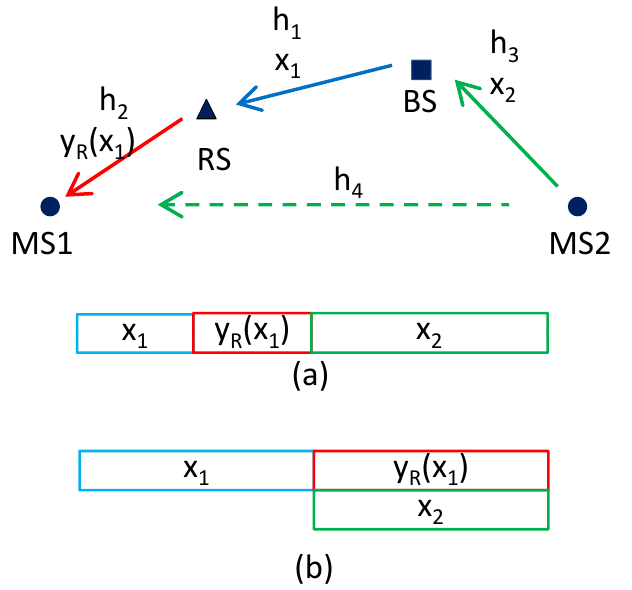}
\caption{Reference scheme 1 (a) and coordinated scheme 1 (b).}
\label{ref1p1}
\end{figure}

The basic setup for a CDR scheme is the scenario with one base station (\verb"BS"), one relay (\verb"RS"), and two users 1 and 2 (\verb"MS1" and \verb"MS2"), see Fig.~\ref{ref1p1}. All transmissions have a unit power and normalized bandwidth of 1 Hz. Each of the complex channels $h_i, i=\overline{1,5}$, is reciprocal, known at the receiver and Rayleigh--faded with parameter $\sigma = 1/\sqrt{2}$. We use the following notation, with a slight abuse: $x_i$ may denote a packet or a single symbol, and it will be clear from the context. For example, the packet that \verb"BS" wants to send to the \verb"MS1" is denoted by $x_1$; but if we want to express the signal received, then we use expressions of type $y=hx_1+z$, where all variables denote symbols (received, sent, or noise). We introduce further notation: $x_4$ is the packet sent from \verb"BS" to \verb"MS2", while the packets that \verb"BS" needs to receive are $x_3$ from \verb"MS1" and $x_2$ from \verb"MS2". Note that the example on Fig.~\ref{ref1p1} does not show traffic patterns that involve $x_3$ and $x_4$.

The basic time unit is one time slot. A direct transmission takes one slot. One transmission through the relay takes also one slot: in the downlink, the first half of the slot is for the transmission \verb"BS-RS", while the second slot is is for the transmission \verb"RS-MS". The uplink transmission is similar.  Relaying with amplify--and--forward (AF) is used, and therefore the transmission \verb"BS-RS" has the same duration with the transmission \verb"RS-MS" (and vice versa in the uplink). The received signal and Additive White Gaussian Noise (AWGN) at \verb"BS", \verb"RS", \verb"MS1" and \verb"MS2" in time slot $j$ is denoted by $y_{ij}$ and $z_{ij}\sim\mathcal{CN}(0, n), ~ i=\{B,R,1,2\}, j=\{1,2\}$. The instantaneous Signal-to-Noise Ratio (SNR) for the $i-$th channel is $\gamma_i=|h_i|^2/n$ and its capacity is denoted as $\C(\gamma_i) = \log_2(1 + \gamma_i)$. The direct channel \verb"BS-MS1" is assumed weak and \verb"MS1" relies only on the amplified/forwarded signal from \verb"RS" in order to decode the signal from \verb"BS". At the RS, the received signal is scaled to comply with transmit constraint.

In the scenarios with more than two users, where scheduling also needs to be applied, there are $k$ relayed users and $k$ direct users. The transmissions are organized in \emph{sessions}. In a session, each user has a packet for an uplink or downlink with probability of $p_u$ or $1 - p_u$, respectively. An example of traffic pattern is shown in table \ref{table} which lists the user numbers in 4 traffic types: direct downlink (DD), direct uplink (DU), relayed downlink (RD) and relayed uplink (RU). Direct users 1, 4, 5 request a downlink, direct users 2, 3 request an uplink, relayed users 6, 7, 9 request a downlink and relayed users 8, 10 request an uplink. One session consists of multiple \emph{frames}, each frame consists of two slots in which a CDR transmission is performed. We slightly abuse the notation for the wireless channels by not explicitly indexing the channel with the particular user: $h_i, i=\overline{1,5}$ thus refers to a channel in the set currently considered for a certain coordination. All channels are assumed known at \verb"BS" and constant during each frame, but vary independently from frame to frame.

\begin{table}[t]
\renewcommand{\arraystretch}{1.3}
\caption{Waiting List}
\label{table}
\centering
\begin{tabular}{|c|c|c|c|}
\hline
 Frame & \textbf{1} & \textbf{2} & \textbf{3} 	\\ \hline
 DD 	& 1 & 4 & 5	\\ \hline
 DU 	& 2 & 3 &  		\\ \hline
 RD 	& 6 & 7 & 9	\\ \hline
 RU 	& 8 & 10 &  	\\ \hline
\end{tabular}
\vspace{-12pt}
\end{table}

\section{Scheduling in Two-User Schemes} \label{two-user schemes}

\begin{figure*}
\centering
\includegraphics[width=1.6\columnwidth]{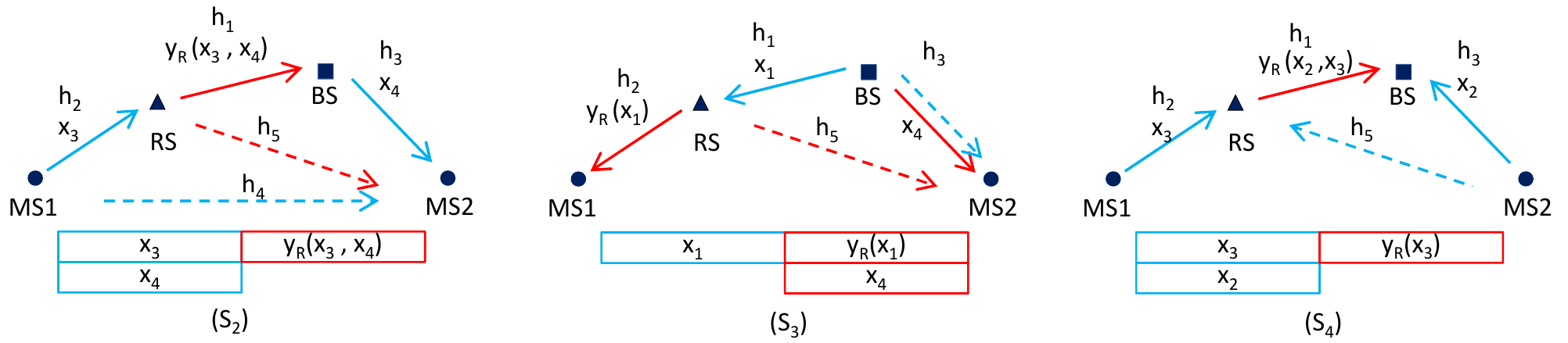}
\caption{Basic coordinated schemes S2, S3 and S4.}
\label{p234}
\end{figure*}

We propose four types of two-user schemes, each combining user pairs (DU, RD), (DD, RU), (DD, RD) and (DU, RU) for which there are packets to be transmitted. These schemes are compared with reference ones in terms of sum--rate. 

\subsection{Reference Schemes}

In the reference schemes there are only orthogonal transmissions and no interference. Four reference schemes corresponding to four user pairs described above have the same time slot structure, only the order and direction of transmissions for the relayed user are different. The reference scheme are denoted $Ei$, $i = \overline{1,4}$.

Fig. \ref{ref1p1} (a) describes the transmissions in reference scheme 1. In the first half slot, \verb"BS" transmits $x_1$ and \verb"RS" receives $y_{R1} = h_1x_1 + z_{R1}$, in the second half-slot, \verb"RS" amplifies $y_{R1}$ with amplification factor $g_{E1} = \frac{1}{|h_1|^2 + n}$ and transmits $\sqrt{g_{R1}}y_{R1}$ and \verb"MS1" receives $y_{11} = h_2\sqrt{g_{R1}}y_{R1} + z_1 = h_2\sqrt{g_{R1}} h_1x_1 + h_2\sqrt{g_{R1}}z_{R1}+ z_1$ , in the second slot, \verb"BS" receives $y_{B2} = h_3x_2 + z_B$. \verb"MS1" decodes $x_1$ from $y_{11}$. SINR for the first user in the reference scheme 1 is therefore
\begin{equation}\gamma_{E11} = \frac{g_{R1}|h_1h_2|^2}{g_{R1}|h_2|^2n + n} = \frac{\gamma_1\gamma_2}{\gamma_1 + \gamma_2 + 1}.\end{equation}
\verb"BS" decodes $x_2$ from $y_{B2}$. SNR for the second user in the reference scheme 1 is $\gamma_{E12} = \frac{|h_3|^2}{n} = \gamma_3$. The scheme thus has sum--rate of $C_{E1} = \frac12\C(\gamma_{E11}) + \C(\gamma_{E12}).$
All reference schemes have the same sum--rate formula due to channel reciprocity and symmetry in AF relaying $C_{E1} = C_{E2} = C_{E3} = C_{E4}.$

\subsection{Proposed Basic Coordinated Schemes} \label{coordinated schemes}
In CDR schemes, the three transmissions are scheduled as one transmission in one slot and two simultaneous transmissions in the other slot although the order and direction of the transmissions are different. The transmissions are arranged so that the interference is reduced or canceled. There are four basic coordinated schemes denoted as S$_i, ~i = \overline{1,4}$. 

\medskip

\noindent \textbf{Coordinated Scheme $S_1$} (Fig. \ref{ref1p1}b), \verb"BS" transmits $x_1$ to \verb"RS" in the first slot, \verb"RS" receives $y_{R1} = h_1x_1 + z_{R1}$. In the second slot, \verb"RS" scales the received signal with the amplification factor $g_{S1} = \frac{1}{|h_1|^2 + n}$ so that the transmit power is 1 and transmit it. At the same time, \verb"MS2" transmits $x_2$. \verb"MS1" therefore receives signal $y_{12} = h_2\sqrt{g_{S1}}y_{R1} + h_4x_2 + z_1 = h_2\sqrt{g_{S1}}h_1x_1 + h_2\sqrt{g_{S1}}z_{R1} + h_4x_2 + z_1$ and \verb"BS" receives $y_{B2} = h_1\sqrt{g_1}y_{R1} + h_3x_2 + z_B =h_1\sqrt{g_{S1}}h_1x_1 + h_1\sqrt{g_{S1}}z_{R1} + h_3x_2 + z_B$. Since \verb"BS" \emph{knows} $x_1$ and the channels, it cancels the component in $x_1$ in $y_{B2}$, gets $\tilde{y}_{B2} = h_3x_2 + h_1\sqrt{g_{S1}}z_{R1} + z_B$ and decodes $x_2$ with SNR
\begin{equation}\gamma_{S12} = \frac{|h_3|^2}{|h_1|^2g_{S1}n + n} = \frac{|h_3|^2(|h_1|^2 + n)}{2|h_1|^2n + n^2} = \frac{\gamma_3(\gamma_1 + 1)}{2\gamma_1 + 1}.\end{equation}
 \verb"MS1" decodes $x_1$ treating $x_2$ as interference with SINR
\begin{equation}\gamma_{S11} = \frac{|h_2|^2g_{S1}|h_1|^2}{|h_2|^2g_{S1} + |h_4|^2 + 1} = \frac{\gamma_1\gamma_2}{\gamma_1+\gamma_2+\gamma_4+\gamma_1\gamma_4+1}.\end{equation}
The sum--rate is therefore $C_{S1} = \C(\gamma_{S11})+\C(\gamma_{S12}).$

\medskip

\noindent \textbf{Coordinated Scheme $S_2$} (Fig. \ref{p234}): \verb"MS1" transmits $x_3$ and \verb"BS" transmits $x_4$ simultaneously in the first slot. \verb"RS" receives $y_{R1} = h_2x_3 + h_1x_4 + z_{R1}$ and \verb"MS2" receives $y_{21} = h_4x_3 + h_3x_4 + z_{21}$. In the second slot, \verb"RS" scales the received signal with the amplification factor $g_{S2} = \frac{1}{|h_1|^2 + |h_2|^2 + n}$ and transmits it. \verb"RS" receives $y_{B2} = h_1\sqrt{g_2}y_{R1} + z_B$ and \verb"MS2" receives $y_{22} = h_5\sqrt{g_{S2}}y_{R1} + z_{22}$. Since \verb"BS" \emph{knows} $x_4$ and the channels, it cancels $x_3$ component in $y_B$, gets $\tilde{y}_{B2} = h_1\sqrt{g_2}(h_2x_3+z_{R1})+z_B$ and decodes $x_3$. At \verb"MS2", $y_{21}$ and $y_{22}$ form a virtual 2-antenna received signal $\mathbf{y} = \mathbf{H}\mathbf{x} + \mathbf{z}$, with $\mathbf{y} = [y_{21} ~y_{22}]^T, \mathbf{x} = [x_4  ~x_3]^T, \mathbf{z} = [z_{21} ~h_5\sqrt{g_{S2}}z_{R1} + z_{22}]^T,$ and
\begin{equation}\mathbf{H} = \left[\begin{array}{cc}h_3 & h_4 \\ \sqrt{g_{S2}}h_1h_5 & \sqrt{g_{S2}}h_2h_5 \end{array} \right].\end{equation}
We can apply MMSE receiver using (\ref{snrmmse}) in the Appendix to have the sum--rate
\begin{equation}C_{S2} = \C\left(\frac{\gamma_1\gamma_2}{2\gamma_1 + \gamma_2 + 1}\right) + C_{S22}\end{equation} in which
$C_{S22} = \C\left[\frac{\gamma_3(\gamma_1 + \gamma_2 + \gamma_5 + 1) + \gamma_5(\gamma_1 + \gamma_{b2})}{(\gamma_4 + 1)(\gamma_1 + \gamma_2 + \gamma_5 + 1) + \gamma_2\gamma_5}\right]$.

\medskip

\noindent \textbf{Coordinated Scheme $S_3$} (Fig. \ref{p234}): \verb"BS" transmits $x_1$ in the first slot, \verb"RS" relays it to \verb"MS1" and \verb"BS" transmits $x_4$ simultaneously in the second slot. The transmissions are $y_{R1} = h_1x_1 + z_{R1}, ~g_{S3} = \frac{1}{|h_1|^2 + n}, ~y_{12} = h_2\sqrt{g_{S3}}y_{R1} + z_{1}, ~y_{21} = h_3x_1 + z_{21}, ~y_{22} = h_5\sqrt{g_{S3}}y_{R1} + h_3x_4 + z_{22}$. \verb"MS1" decodes $x_1$ from $y_{12}$ without interference. At \verb"MS2", $y_{21}$ and $y_{22}$ form a virtual 2-antenna received signal $\mathbf{y} = \mathbf{H}\mathbf{x} + \mathbf{z}$, with $\mathbf{y} = [y_{21} ~y_{22}]^T, \mathbf{x} = [x_4  ~x_1]^T, \mathbf{z} = [z_{21} ~h_5\sqrt{g_{S3}}z_{R1} + z_{22}]^T,$ and
$\mathbf{H} = \left[\begin{array}{cc}0 & h_3 \\ h_3 & \sqrt{g_{S3}}h_1h_5 \end{array} \right]$.
We can apply MMSE receiver using (\ref{snrmmse}) in the Appendix to have the sum--rate
\begin{equation}C_{S3} = C_{S31} + C_{S32} = \C\left(\frac{\gamma_1\gamma_2}{\gamma_1 + \gamma_2 + 1}\right) + C_{S32}\label{cp3}\end{equation}
with $C_{S32} = \C\left[\frac{\gamma_3(\gamma_3 + 1)(\gamma_1 + 1)}{(\gamma_1 + \gamma_5 + 1)(\gamma_3 + 1) + \gamma_1\gamma_5}\right]$.

\medskip

\noindent \textbf{Coordinated Scheme $S_4$} (Fig. \ref{p234}): \verb"MS1" transmits $x_3$ and \verb"MS2" transmits $x_2$ in the first slot, \verb"RS" transmits what it received in the second slot. The transmissions are $y_{R1} = h_2x_3 + h_5x_4 + z_{R1}, ~g_{S4} = \frac{1}{|h_2|^2 +|h_5|^2 + n}, ~y_{B1} = h_3x_4 + z_{B1}, ~y_{B2} = h_1\sqrt{g_{S4}}y_{R1} + z_{B2}$. \verb"BS" decodes $x_2$ and $x_3$ from $y_{B1}$ and $y_{B2}$. Similar to the previous schemes, we have $\mathbf{y} = \mathbf{H}\mathbf{x} + \mathbf{z}$, with $\mathbf{y} = [y_{B1} ~y_{B2}]^T, \mathbf{x} = [x_2  ~x_3]^T, \mathbf{z} = [z_{B1} ~h_1\sqrt{g_{S4}}z_{R1} + z_{B2}]^T,$ and
$\mathbf{H} = \left[\begin{array}{cc}0 & h_3 \\ \sqrt{g_{S4}}h_1h_2 & \sqrt{g_{S4}}h_1h_5 \end{array} \right]$.
We can apply MMSE receiver using (\ref{snrmmse}) in the Appendix for both users to have the sum--rate
we have $C_{S4} = C_{S41} + C_{S42}$, in which
\begin{equation}C_{S41} = \C\left[\frac{\gamma_1\gamma_2(\gamma_3 +1)}{\gamma_1 \gamma_5  + (\gamma_1 + \gamma_2 + \gamma_5 + 1)(\gamma_3 +1)}\right],\end{equation}
where 
\begin{equation}
C_{S42} = \C\left[\gamma_3 + \frac{\gamma_1\gamma_5}{\gamma_1 + \gamma_2 + \gamma_5 + \gamma_1\gamma_2 + 1}\right]
\label{cp42}
\end{equation}

\subsection{Coordinated Schemes with User Priority}
In the reference schemes, information for the relayed and direct user is transmitted separately. On average, the BS-relayed user rate is lower than the BS-direct user rate because of AF relaying. However, in some of the coordinated schemes, the BS-relayed user rate may be higher than or approximately equal to the BS-direct user rate. This can be unfair for the direct user which deserves a higher rate. In addition, with the same amount of resource (transmit power, time slots...) a coordinated scheme can increase the sum-rate by allocating more resource to the direct user than the amount which is used in basic coordinated schemes. In this section, we introduce a proritizing factor denoted as $\lambda \in (-1, 1)$. If $\lambda > 0$, the direct user has more priority than in the basic coordinated scheme and the relayed user has less priority and vice versa.

\begin{figure}
\centering
\includegraphics[width=0.9\columnwidth]{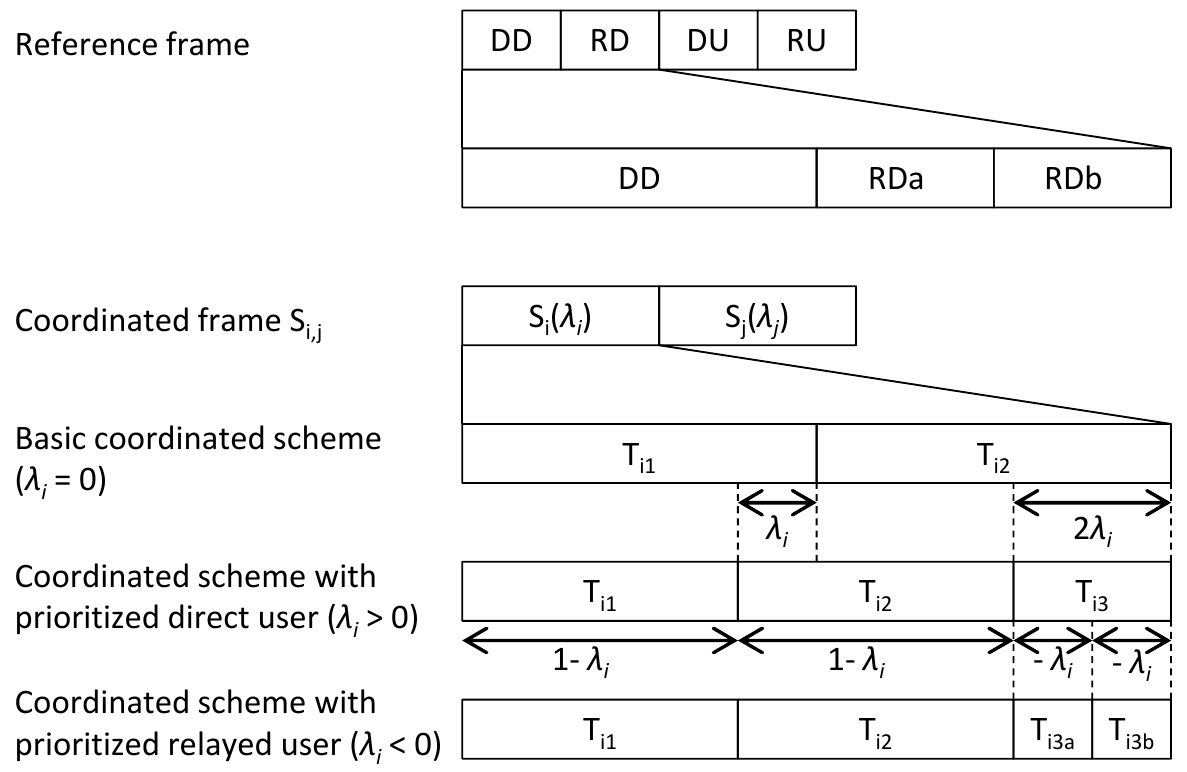}
\caption{Frame format for reference, basic and prioritizing coordinated schemes. a or b refers to either of the BS-RS or RS-relayed user transmission.}
\label{frame structure}
\end{figure}

When $\lambda >0$, the scheme begins with two time slots as in a basic coordinated scheme however the length of each time slot is $1 - \lambda$ symbol time instead of 1 symbol. The residual time with a length of $2\lambda$ is used for an additional transmission between BS and the direct user (Fig. \ref{frame structure}). When $\lambda < 0$, the additional transmissions are from BS to RS and RS to the relayed user for a relayed downlink or in opposite order for a relayed uplink. If $\lambda = 1$ or $\lambda = -1$, the whole time is used for the direct user or the relayed user respectively, which is not considered.

The average rate for coordinated scheme $i$ with prioritizing factor $\lambda$ is
\begin{equation}
C'_{Si} = \left\{ \begin{array}{rcl} (1 - \lambda)(C_{Si1} + C_{Si2}) + 2\lambda\C(\gamma_3) & \mbox{for} & \lambda > 0 \\
(1 + \lambda)(C_{Si1} + C_{Si2}) - \lambda\C(\gamma_R) & \mbox{for} & \lambda < 0 \end{array}\right.
\end{equation}
with $\gamma_R = \frac{\gamma_1\gamma_2}{\gamma_1 + \gamma_2 +1}$.

\section{Scheduling in Multi-User Schemes} \label{multi-user schemes}
This part presents different ways of scheduling transmissions in a session. At the beginning of a session, \verb"BS" receives requests for uplink/downlink and channel information from some users. It schedules appropriate transmissions in the first frame according to one of the schemes below. After that, it receives channel information and schedules for the new frame and so on until all requests in the session are fulfilled.

Note that one frame has two coordinated schemes, such that we use the notation $S_{i,j}$ to denote the fact that the frame contains the coordinated schemes $S_i$ and $S_j$.

\subsection{Multi-user Reference Scheme}
A frame in the multi-user reference scheme contains 4 time slots each for 4 traffic types (DD, RD, DU, RU). The users in each traffic type are served according to first-in-first-out discipline. If there is not a packet corresponding to a slot, it is left empty. A slot for a relayed user is divided into two small slots: one for the transmission between \verb"BS" and \verb"RS" and one for the transmission between \verb"RS" and the relayed user. In the example of table \ref{table}, the first, second and third frames are (1, 2, 6, 8), (4, 3, 7, 10) and (5, $\varnothing $, 9, $\varnothing $) respectively. In slot $\varnothing $ there is no transmission.

\subsection{Coordinated Schemes}
Instead of transmitting packets of 4 traffic types separately in a frame, we can use the coordinated schemes described in part \ref{coordinated schemes} to combine the transmissions. Two types of combining are S$_{1,2}$ which includes S$_1$ (DU, RD) and S$_2$ (DD, RU) and S$_{3,4}$, which includes S$_3$ (DD, RD) and S$_4$ (DU, RU). The packets are thus transmitted frame by frame according to first-in-first-out discipline using one of these two schemes. If S$_{3,4}$ is used in the example in table \ref{table}, the frames are [(1, 6), (2, 8)], [(4, 7), (3, 10)], [(5, 9), $\varnothing$].

\subsection{Proposed Multi-user Schemes}
As users experience different channel qualities within each frame, the achieved performance will highly depend on the chosen user combinations, in each scheme. With an exhaustive search among the users and sum--rate estimating for each case and both schemes S$_{1,2}$ and S$_{3,4}$, we can find a combination of four users and a scheme which has the highest sum--rate in the current frame.

The complexity for CDR with exhaustive search, however, is prohibitively high since we have to calculate the sum--rate for every permutation of the packets. It is necessary to propose some sub-optimal schemes which requires a lower complexity without a significant rate loss. Such a reduction of the search space would be, for example, if only combination of $S_3$ and $S_4$ in a frame is considered (thus, the $S_{3,4}$ schemes), without considering the possibility to use the schemes $S_1$ and $S_2$. 

Another suboptimal scheme is Best Direct User CDR (BDCDR): first, the direct downlink user which has the best channel to \verb"BS" ($\max(\gamma_3)$) is picked. The downlink relayed user which has the best combination with that direct user is chosen after that. Then we have a sub-optimal coordinated scheme of S3. We can do similarly for S1, S2, S4 and choose the higher of S12 and S34. In Best Relayed User CDR scheme (BRCDR), the relayed downlink user which has the best relayed channel to \verb"BS" ($\max(\frac{\gamma_1\gamma_2}{\gamma_1 + \gamma_2 + 1})$) picked first. The steps after that are processed in a similar way to BDCDR.

\section{Numerical Results} \label{numerical results}
\begin{figure}
\centering
\includegraphics[width=0.8\columnwidth]{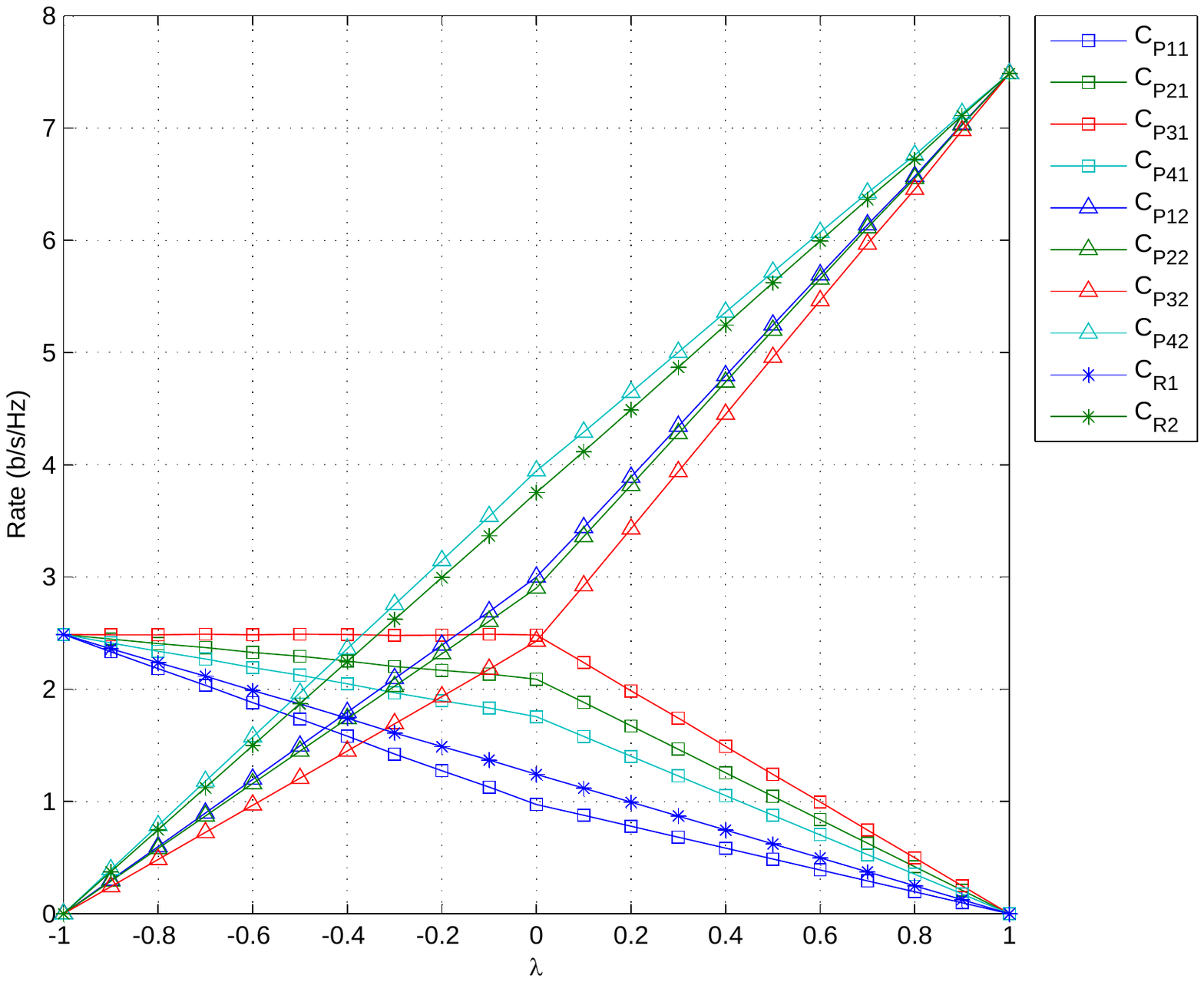}
\caption{Average rates for different prioritizing factor values}
\label{lambda}
\end{figure}
Computer simulations with network scenarios and parameters as presented in part \ref{system model} is conducted to illustrate the sum–rate for the reference and proposed schemes.

In case of two users, Fig. \ref{lambda} shows the rate for the relayed user (C$_{i1}$, $i \in \{E, C1, C2, C3, C4 \}$), and the direct user (C$_{i2}$) changing as a function of prioritizing factor in reference and coordinated schemes. The case when $\lambda = 0$ is correspondent to the basic reference and coordinated schemes. When $\lambda$ increases, the direct user become more prioritized and in the opposite direction the relayed user does. Prioritizing the direct user gives more improvement than prioritizing the relayed user in terms of sum--rate because AF relaying communication between the relayed user and BS. $C_{S42}$ is the highest among $C_{i2}$ because as seen in (\ref{cp42}), always higher than $C_{E2}=\C(\gamma_3)$, that \verb"BS" exploits the information in the second time slot beside the information it receives in the first time slot which is equal to $\C(\gamma_3)$. $C_{S32}$, always lower than $\C(\gamma_3)$, is the lowest among $C_{i2}$ because \verb"MS2" only receives information in time slot 2 which is interfered by the transmission of \verb"RS". On contrary, $C_{S31}$ is the highest among $C_{i1}$ since \verb"MS1" receives information from \verb"RS" without any interference in a full time slot compared to half slot in reference scheme. Because $\gamma_{S31} = \gamma_R$, when $\lambda < 0$, $C_{S31}$ does not change therefore decreasing a negative $\lambda$ does not bring any benefit but a decrease in the sum--rate. $C_{S11}$ is the lowest due to the interference from \verb"MS2" at \verb"MS1" over the inter-user channel.

\begin{figure}
\centering
\includegraphics[width= 0.8\columnwidth]{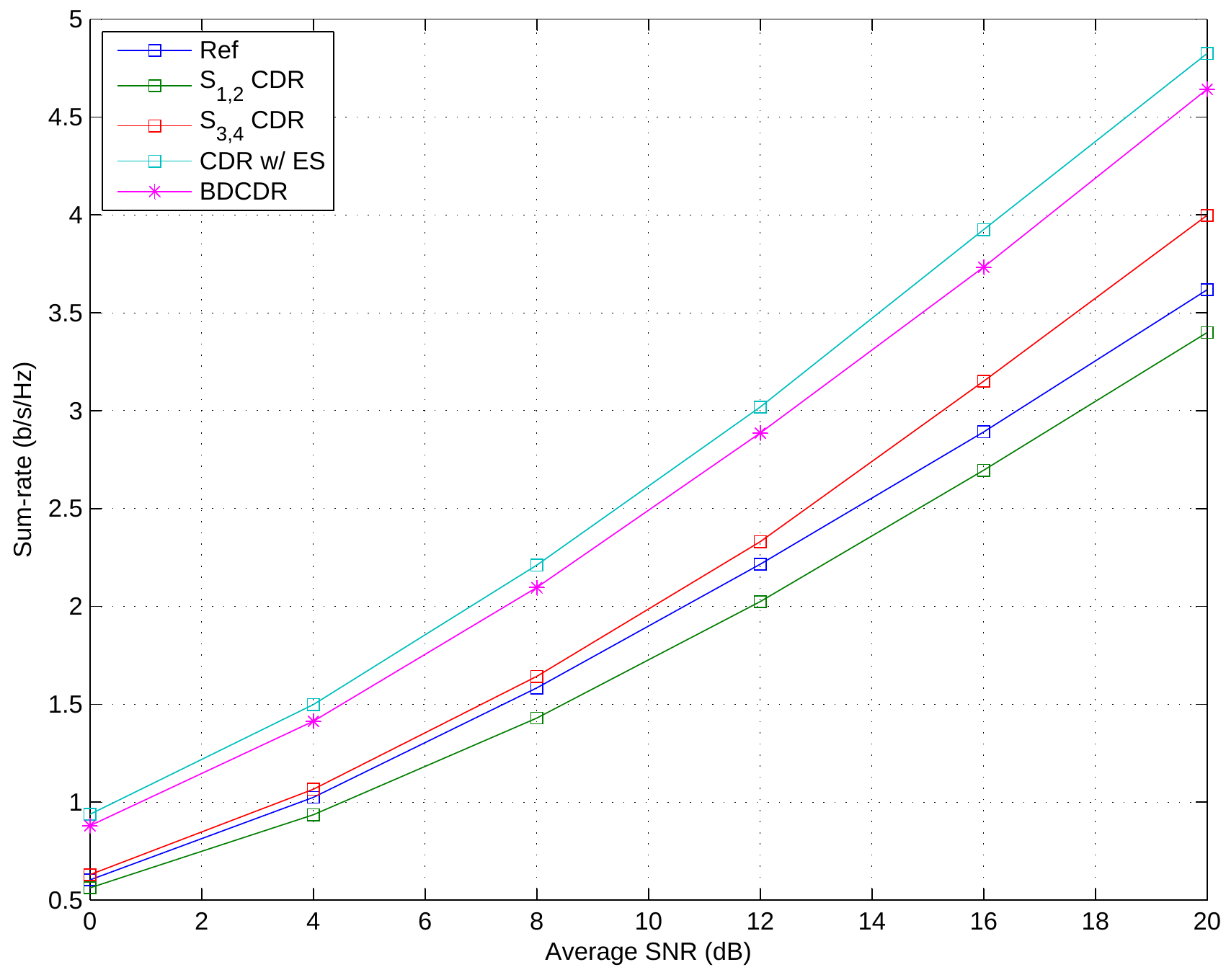}
\caption{Average rates for different ordering schemes}
\label{j1}
\end{figure}

In multi-user case with $k = 10$, Fig. \ref{j1} compares the sum--rate for different CDR schemes with $p_u = \frac12, \lambda = 0$. The CRD scheme with exhaustive search has the highest sum--rate due to the optimal combination of transmissions in each frame. The BDCRD scheme with much lower complexity achieves a slightly lower sum--rate since the rate for all scheme increases with $\gamma_3$ which is optimal in a Best Direct scheme and the channels are independently distributed. CDR with only S$_{3,4}$ has much lower sum--rate since there is no choice among the users. The S$_{1,2}$ CDR even has a lower sum--rate than the reference scheme due to useless inter-user interference in S$_1$.

\section{Conclusion} \label{conclusion}
In this paper, we have proposed  coordinated schemes for a network with a direct user, a relayed user, a base station and a relay station. These schemes are inspired by the principles used in physical--layer network coding: uplink and downlink flows are aggregated, while the interference is not avoided by orthogonalization, but rather through the usage of information that is known \emph{a priori}. The proposed schemes are shown to have higher sum--rate than the conventional schemes. These schemes 
are then used as building blocks in multi--user scenarios, where we present several 
schemes for scheduling pairs of users for coordinated transmissions. In order to avoid scheduling complexity, we propose several suboptimal scheme, which perform closely to the optimal scheme. We have also discussed the trade-off between the network sum--rate and user prioritization. As a future work, we intend to analyze the proposed schemes in frameworks with proportional fair scheduling and propose related schemes in multi--channel systems, such as OFDMA. 

\appendix
Consider a 2x2 MIMO system with transmit vector $\mathbf{x} = [x_1~x_2]^T$, receive vector $\mathbf{y}$ and channel matrix $\mathbf{H}$. We have transmission vector equation
\setlength{\arraycolsep}{0.5em}
\begin{equation}\mathbf{y}=\mathbf{H}\mathbf{x}+\mathbf{z} = \mathbf{h_1}x_1 + \mathbf{h_2}x_2 + \mathbf{z}.\end{equation}
Denote
\begin{equation}\mathbf{H} =\left[\begin{array}{cc}\mathbf{h_1} & \mathbf{h_2}  \end{array}\right] =  \left[\begin{array}{cc}h_{11} & h_{21} \\ h_{12} & h_{22} \end{array} \right],\end{equation}
\begin{equation}\mathbf{\Gamma} = \left[\begin{array}{cc}\gamma_{11} & \gamma_{21} \\ \gamma_{12} & \gamma_{22} \end{array} \right] = \frac{1}{n}\left[\begin{array}{cc}\ |h_{11}|^2 & |h_{21}|^2 \\ |h_{12}|^2 & |h_{22}|^2 \end{array} \right],\end{equation} 
\begin{align*}\gamma_a &= \frac{|h_{11}^*h_{21} + h_{12}^*h_{22}|^2}{n^2}, &\gamma_b &= \frac{|h_{11}h_{22} - h_{21}h_{12}|^2}{n^2},\end{align*}
\setlength{\arraycolsep}{0.5em}
\begin{equation}\mathbf{N} = \E[\mathbf{z} \mathbf{z}^H]= n\left[\begin{array}{cc}1 & 0\\ 0 & \alpha  \end{array} \right].\end{equation}

By noise whitening \cite{tse}, we have
\begin{equation}\mathbf{K}_z = \E[(\mathbf{h_2}x_2 + \mathbf{z})(\mathbf{h_2}x_2 + \mathbf{z})^H] = \mathbf{h_2}\mathbf{h_2}^H + \mathbf{N}\end{equation}
\begin{equation}\mathbf{h_1}^H\mathbf{K}_z^{-1}\mathbf{y} = \mathbf{h_1}^H\mathbf{K}_z^{-1}\mathbf{h_1}\mathbf{x} + \mathbf{h_1}^H\mathbf{K}_z^{-1}(\mathbf{h_2}x_2 + \mathbf{z}).\end{equation}
SINR matrix is therefore \begin{equation}\mathbf{SINR}=\mathbf{h_1}^H\mathbf{K}_z^{-1}\mathbf{h_1}\end{equation}
in which SINR for the first information stream is\begin{equation}\label{snrmmse1}SINR_1=\frac{\alpha\gamma_{11} + \gamma_{12} + \gamma_b}{\alpha \gamma_{21} +  \gamma_{22} + \alpha}\end{equation}
In case the noise power matrix is in another form
\begin{equation}\E[\mathbf{z}\mathbf{z}^H]= n\left[\begin{array}{cc}\alpha & 0\\ 0 & 1  \end{array} \right],\end{equation}
we have \begin{equation}\label{snrmmse}SINR_1=\frac{\gamma_{11} + \alpha\gamma_{12} + \gamma_b}{ \gamma_{21} +  \alpha\gamma_{22} + \alpha}.\end{equation}

\setlength{\arraycolsep}{5em}

\section*{Acknowledgment}
This work is supported by the Danish Research Council for
Technology and Production, grant nr. $09-065035$.

\end{document}